\documentclass[twocolumn]{aastex63}

\usepackage{amsmath}

\shorttitle{Evolutionary Origin of Ultra-long Period Radio Transients}
\shortauthors{Fan et al.}

\begin{document}


\title{Evolutionary Origin of Ultra-long Period Radio Transients}


\author[0009-0006-0116-5175]{Yun-Ning Fan}
 \affil{School of Science, Qingdao University of Technology, Qingdao 266525, People's Republic of China; chenwc@pku.edu.cn}
\author[0000-0002-9739-8929]{Kun Xu}
 \affil{School of Science, Qingdao University of Technology, Qingdao 266525, People's Republic of China; chenwc@pku.edu.cn}
\author[0000-0002-0785-5349]{Wen-Cong Chen}
  \affil{School of Science, Qingdao University of Technology, Qingdao 266525, People's Republic of China; chenwc@pku.edu.cn}
  \affil{School of Physics and Electrical Information, Shangqiu Normal University, Shangqiu 476000, People's Republic of China}



\begin{abstract}
Recently, it discovered two ultra-long period radio transients GLEAM-X J162759.5-523504.3 (J1627) and GPM J1839$-$10 (J1839)
with spin periods longer than 1000 s. The origin of these two ultra-long period radio transients is intriguing in understanding
the spin evolution of neutron stars (NSs). In this work, we diagnose whether the interaction between strong magnetized NSs and
fallback disks can spin NSs down to the observed ultra-long period. Our simulations found that the magnetar+fallback disk model can account for the observed period, period derivative, and X-ray luminosity of J1627 in the quasi-spin-equilibrium stage. To evolve to the current state of J1627, the initial mass-accretion rate of the fallback disk and the magnetic field of the NS are in the range of $(1.1-30)\times10^{24}~\rm g\,s^{-1}$ and $(2-5)\times10^{14}~\rm G$, respectively. In an active lifetime of fallback disk, J1839 is impossible to achieve the observed upper limit of period derivative. Therefore, we propose that J1839 may be in the second ejector phase after the fallback disk becomes inactive. Those NSs with a magnetic field of $(2-6)\times10^{14}~\rm G$ and a fallback disk with an initial mass-accretion rate of $\sim10^{24}-10^{26}~\rm g\,s^{-1}$ are the possible progenitors of J1839.
\end{abstract}


\section{Introduction}
Since the first radio pulsar PSR B1919+21 was discovered \citep{hewi68}, about 3500 pulsars have
been discovered up to now \citep{manc05}. Radio pulsars are generally thought to be rapidly
rotating strongly magnetized neutron stars (NSs), in which the radio emission is powered by the rotational
kinetic energy via magnetic dipole radiation \citep{gold68}. In the spin period ($P$) versus spin-period derivative
($\dot{P}$) diagram, the $P$ and $\dot{P}$ of normal radio pulsars are $\sim1$ s and $\sim10^{-15}~\rm s\,s^{-1}$ \citep{taur06},
respectively. As a special population of NSs, the magnetars possess strong magnetic fields ($\sim10^{14}~\rm G$) and long spin periods in the range of $2 - 12~\rm s $ \citep{olau14}. The historical lack of radio pulsars with periods longer than 12 s was interpreted to be the existence of the "death line", under which the voltage generated above the polar cap is below a critical value \citep{rude75,bhat91,chen93}.

The discovery of some long-period pulsars challenges the conventional pulsar death line model. PSR J0250+5854 (J0250) is a radio
pulsar with a spin period of 23.5 s, which is in the conventional pulsar death line \citep{tan18}. The Galactic Plane Pulsar Snapshot survey also discovered a long-period pulsar PSR J1903+0433g with a spin period of 14.05 s \citep{han21}. The discovery of GLEAM-X J162759.5-523504.3 (hereafter J1627) has excited the pulsar community and marked the start of a new era of ultra-long period pulsar field. \cite{hurl22} detected a periodic, low-frequency radio transient J1627, which pulses every 18.18 minutes (1091s). This unusual periodicity challenges our knowledge of the radiation mechanism of pulsars. Subsequently, the discovery of PSR J0901-4046 (J0901) with a spin period of 75.88 s supplied a new member for the long-period pulsar population \citep{cale22}. Recently, the discovery of an ultra-long period radio transient GPM J1839$-$10 (hereafter J1839) is icing on the cake in this population. J1839 was detected to be repeating at a pulsation period of 1318 s since at least 1988, and its period derivative is constrained to be less than $3.6\times10^{-13}~\rm s\,s^{-1}$ \citep{hurl23}.

It is weird that the radio luminosity ($4\times10^{31}~\rm erg\,s^{-1}$) inferred from the brightest pulses of J1627 is much higher than its maximum spin-down power \cite[$\dot{E}=1.2\times10^{28}~\rm erg\,s^{-1}$,][]{hurl22}.
Therefore, \cite{hurl22} proposed that J1627 is a radio magnetar rather than a pulsar according to the properties including the smooth variations in pulse profile and the transient window of radio emission \citep{levi12}. Nevertheless, the analysis based on the spectral data suggested that the highest possible radio luminosity of J1627 does not exceed its spin-down luminosity \citep{erku22}. Because of long spin periods, long-period pulsars were proposed to be most likely white dwarfs \citep{katz22} or hot subdwarfs
\citep[proto white dwarf,][]{loeb22}. However, \cite{beni23} demonstrated that the observations of J1627 are unlikely to be explained by either a magnetically or a rotationally powered white dwarf. \cite{kou19} proposed that the long period of J0250 is related to the
magnetospheric evolution and magnetic field decay. \cite{ronc22} showed that the newly born NSs with strong
magnetic fields of $10^{14}-10^{15}$ G and fallback disks with initial accretion rates of $10^{22}-10^{27}~\rm g\,s^{-1}$
could evolve into long-period isolated radio pulsars like J1627 and J0901 in a short timescale of $10^{3}-10^{5}~\rm yr$.

The first candidate of long-period magnetars is the NS 1E 161348$-$5055 in the supernova remnant RCW 103,
which has an incredibly long period of 6.67 hours \citep{luca06}. Many features of this NS are similar to those of traditional
magnetars such as anomalous X-ray pulsars and soft gamma-ray repeaters, including a magnetar-like short X-ray burst \citep{dai16},
longer-term outbursts, and a hard X-ray tail in the spectrum during outburst \citep{rea16}. The NS should be a young magnetar because the age of RCW 103 was inferred to be $\sim3.3$ kyr \citep{clar76} or in the range of $1.2-3.2$ kyr \citep{nuge84,cart97}.
Recently, detailed modeling of the supernova remnant estimated the age of the source to be $880-4400$ yr \citep{brau19}. The emission properties also imply that 1E 161348$-$5055 should be a young magnetar, in which the 6.67 hr periodicity can only be
thought to be the spin period of the magnetar \citep{rea16}. It is extremely anomalous that the spin period of this source is much
longer than those ($2-12$ s) of traditional magnetars. Ignored the ejector phase, \cite{luca06} found that
a magnetic field of $B=5\times10^{15}$ G and a debris disc mass of $3\times10^{-5}~M_{\odot}$ can spin the NS down to the present
period in propeller phase. Subsequent some detailed simulation also confirmed the magnetar nature of 1E 161348$-$5055 \citep{li07,luca08,espo11,dai16,rea16,tong16,ho17,tend17,borg18,xu19}. \cite{beni20} demonstrated that an episodic mass-loaded charged particle winds can efficiently spin a magnetar down to an ultra-long period.

Compared with 1E 161348$-$5055, J1627 and J1839 were detected the upper limit of period derivative as $\dot{P}= 1.2\times 10^{-9}~ \rm s\ s^{-1} $ \citep{hurl22} and $ \dot{P}=3.6\times 10^{-13}~\rm s\,s^{-1}$ \citep{hurl23}, respectively. Especially, J1839 is located at the very edge of an extremely tolerant death line, making it impossible as a classical radio pulsar \citep{hurl23}. However, \cite{tong23a} suggested that the pulsar death line of long-period pulsars should be revised due to two possible physical
effects of a fallback disk or a twisted magnetic field, and J1627 may be a radio-loud magnetar spun down by a fallback disk.
Recently, J1627 and J1839 were thought to be most likely magnetars with twisted magnetic fields, magnetars with fallack
disk, or white dwarf radio pulsars \citep{tong23b}. Therefore, the possibility that J1839 is a radio magnetar cannot be completely excluded.

In this work, we attempt to diagnose whether the magnetar + fallback disk model can account for the long spin periods and
period derivative of ultra-long period radio transients J1627 and J1839. Meanwhile, we also investigate the initial parameter
space producing these two transients. In Section 2, we describe
the detailed physical model including fallback disk, ejector phase, and propeller phases. The simulated results
of the sources J1627 and J1839 are summarized in Section 3. Finally, we give a  brief discussion and summary
in Sections 4 and 5.

\section{Physical Model}
\subsection{Fallback disk}
Isolated NSs are the evolutionary products of massive stars through core-collapse supernovae
when their progenitors exhaust all fuels. During supernovae explosions, a tiny fraction of ejecta
could form a fallback disk surrounding the nascent NS rather than completely leave it because
those ejecta possess a sufficient angular momentum \citep{mich88,lin91,pern14}. Two normal magnetars 4U 0142+61
and 1E 2259+586 are most likely to be surrounded by fallback disks \citep{wang06,kapl09}. Without the
interaction between the fallback disk and the NS, the spin evolution of the NS is driven by
magnetic dipole radiation and is in an ejector phase in the early stage. With the increase of the spin
period, the fallback disk can interact with the NS magnetosphere, and the NS enters a propeller phase.
In this phase, a propeller torque originated
from the interaction between the fallback disk and the NS can significantly influence the spin
evolution of the NS \citep{illa75}, and the spin angular velocity of the NS decreases at an exponential
rate for a short time $t_{\rm prop}$ \citep[see also Eq. \ref{eq:propeller},][]{ho17}.


In the most lifetime, the accretion rate in the fallback disk decreases self-similarly
according to a power-law as $\dot{M}\propto t^{-\alpha}$ ($t$ is the age of the fallback
disk), in which $\alpha=19/16$ for a fallback disk whose opacity is dominated by electron
scattering \citep{cann90}. In the calculation, we adopt an accretion rate whose evolutionary
law same to \cite{chat00} as follows
\begin{equation}
	\dot{M}=\left\{
	\begin{array}{ll}
		\dot{M}_{0}, & t<T \\
		\dot{M}_{0}(t / T)^{-\alpha}, & t \geq T
	\end{array}\right.
\end{equation}
where $\dot{M}_{0}$ is the initial mass-accretion rate when $t<T$, in which $T$ is of the order
of the dynamical timescale in the inner region of the nascent fallback disk.
\cite{chat00} adopted a typical value of $T=1~\rm ms$. Similar to \cite{meno01}, we take $T=2000~\rm s$ \citep{tong16,ronc22},
which can be derived from a typical viscous timescale that the fallback material with excess angular momentum
circularizes to form a disk. Since $T=6.6\times10^{-5}~\rm yr$ is much
smaller than the age ($10^{3}-10^{4}~\rm yr$) of long-period pulsars, the accretion process in the early stage ($t<T$)
is negligible in our simulations. Since the initial mass of the fallback disk satisfies
\begin{equation}
M_{\rm d,i}=\dot{M}_{0}T+\int^{\infty}_{T}\dot{M}_{0}(t / T)^{-\alpha}{\rm d}t,
\end{equation}
we have $\dot{M}_{0}=(\alpha-1)M_{\rm d,i}/(\alpha T)$ \citep{chen22}.

In general, the inner radius of the fallback disk is thought to be the magnetospheric
radius, at which the magnetic energy density is equal to the kinetic energy density
of the inflow material. The magnetospheric radius is given by \citep{davi73,elsn77,ghos79}
\begin{equation}
 \begin{aligned}
{R}_{\rm m}=\xi\left(\frac{\mu^{4}}{2GM\dot{M}^{2}}\right)^{1/7}=7.5\times10^{8} {\rm cm}~\xi_{0.5} \\
\times M_{1.4}^{-1/7}B_{14}^{4/7} R_{6}^{12/7} \dot{M}_{18}^{-2/7},\label{eq:rm}
 \end{aligned}
 \end{equation}
where $G$ is the gravitational constant, $M$ is the NS mass, $\dot{M}$ is the accretion rate,
$\mu=BR^{3}/2$ ($B$ and $R$ are the surface magnetic field and the radius of the NS) is the
the magnetic dipole moment of the NS, and $\xi\approx 0.5$ is a corrective factor \citep{wang96,long05}.
In equation \ref{eq:rm}, $M_{1.4}=M/1.4~M_{\odot}$, $R_{6}=R/10^{6}~\rm cm$, $B_{14}=B / 10^{14}~\rm G $, and
$\dot{M}_{18}=\dot{M} / 10^{18}~ \rm g\,s^{-1}$.

\subsection{Ejector phase}
In the early stage after the NS was born, the inner radius of the fallback disk is greater than
the light-cylinder radius, which is defined as
\begin{equation}
R_{\rm lc}= Pc/2\pi=4.8\times10^{6} {~\rm cm}P_{-3},
\end{equation}
where $c$ is the speed of light in vacuo, $P_{-3}=P/10^{-3}~\rm s$ is the spin period of the NS.
In this stage, the rapidly rotating NS cannot interact with the accreted material, appearing as
a radio pulsar, that is the ejector phase. The NS could radiate strong radio emission by the
magnetic dipole radiation, which exerts a braking
torque on the NS as follows
\begin{equation}
N_{\rm md}=-\frac{2\mu^{2}\Omega^{3}\sin^{2}\theta }{3c^{3}} =-\beta I\Omega^{3},
\end{equation}
where $\Omega=2\pi/P$ is spin angular velocity of the NS, $\theta$ is the inclination angle
between the magnetic axis and the rotation axis, $I$ is the momentum of inertia. For simplicity,
in this work we assume an orthogonal rotator,  i.e. $\theta=\pi/2$, thus $\beta \equiv 2\mu^{2}/3c^{3}I$.
According to the law of rotation $N_{\rm md}=I\dot{\Omega}$, we have $ \dot{\Omega}=-\beta\Omega^{3}$.
Therefore, in the ejector phase, the spin period of the NS satisfies
\begin{equation}
P=P_{0}\left ( 1+\frac{8\beta\pi^{2} t}{P_{0}^{2}}  \right )^{1/2}=
P_{0}\left( 1+\frac{t}{t_{\rm em}}  \right )^{1/2}, \label{eq:ejectorP}
\end{equation}
where $P_{0}$ is the initial spin period of the NS. Due to magnetic dipole radiation, the nascent NS
spin down on the timescale
\begin{equation}
{t_{\rm em}}=\frac{P_{0}}{\dot{2P_{0}}}=\frac{P_{0}^{2}}{8\pi^{2}\beta}=2.1\times10^{5}~{\rm s}~\frac{P_{0,-3}^{2}I_{45}}
{B_{14}^{2}R_{6}^{6}},
\end{equation}
where $I_{45}=I/10^{45}~\rm g\,cm^{2}$.

With the spin-down of the NS, the light-cylinder radius gradually increases. Once $R_{\rm m}<R_{\rm lc}$,
the NS magnetosphere can interact with the fallback disk, and the propeller phase begins. The transition
period (i.e. the maximum period of the ejector phase) between the ejector phase and the propeller phase is
\begin{equation}
\begin{aligned}
	P_{\rm ej,max}=\frac{2\pi}{\Omega_{\rm ej,max}}=\frac{2\pi R_{\rm m}}{c}=0.16~{\rm s}~\xi_{0.5} \\
\times M_{1.4}^{-1/7}B_{14}^{4/7} R_{6}^{12/7} \dot{M}_{18}^{-2/7}. \label{eq:Pejmax}
\end{aligned}
\end{equation}
Inserting Equation (8) into Equation (6), we can estimate the duration of the ejector phase to be
\begin{equation}
\begin{aligned}
	t_{\rm ej}=t_{\rm em}\left[ \left ( \frac{P_{\rm ej,max}}{P_{0}}  \right )^{2}-1 \right]\approx164~
{\rm yr}~\xi_{0.5}^{2} \\
\times M_{1.4}^{-2/7}R_{6}^{-18/7}I_{45}B_{14}^{-6/7}  \dot{M}_{18}^{-4/7}.
\end{aligned}
\end{equation}

In Equation (9), $\dot{M} $ is time-varying, while $ t_{\rm ej}$ should be a constant. According
to $\dot{M}_{18}=\dot{M}_{\rm 0,18}(t/T)^{-19/16}$, $ t=t_{\rm ej}$,
and Equation (8), it yields
\begin{equation}
t_{\rm ej}=\left(164~ {\rm yr}~\xi_{0.5}^{2} M_{1.4}^{-\frac{2}{7}}R_{6}^{-\frac{18}{7}} I_{45}B_{14}^{-\frac{6}{7}}
\dot{M}_{0,18}^{-\frac{4}{7}} T^{-\frac{19}{28}}\right)^{\frac{28}{9}},
\end{equation}
where $T$ in units of second.
\subsection{Propeller phase}
The NS enters the propeller phase once the inner radius of the fallback disk is located between the
light-cylinder radius and the corotation radius, in which the corotation radius is defined as
\begin{equation}
R_{\rm co}=\sqrt[3]{\frac{GMP^{2}}{4\pi^{2}}}=1.7\times10^{6} {~\rm cm}\  M_{1.4}^{1/3}P_{-3}^{2/3}.
\end{equation}
During the propeller phase ($R_{\rm c}\leq R_{\rm m}\leq R_{\rm lc}$), the Keplerian
angular velocity ($\Omega_{\rm K}(R_{\rm m})=\sqrt{GM/R_{\rm m}^{3}}$)  at the magnetosphere radius of the fallback disk
is smaller than the spin angular velocity of the NS. Therefore, the accreted material cannot keep co-rotating with the NS
and was thought to be ejected at $R_{\rm m}$ due to the centrifugal barrier, resulting in a negative
torque exerted on the NS. Meanwhile, the interaction between the magnetic
lines and the fallback disk also produces a similar negative torque \citep{meno99}.
The detailed numerical simulations indicate that the two braking torques mentioned
above is approximately equal \citep{daum96}. As a consequence, a braking torque exerting
on the NS in the propeller phase can be expressed as
\begin{equation}
N_{\rm prop}=2\dot{M}R_{\rm m}^{2}(\Omega_{\rm K}(R_{\rm m})-\Omega),
\end{equation}
where the factor of 2 originates from the two nearly equal negative torques mentioned above.
Our adopted braking torque is two times as large as that taken by \cite{ho17}, in which the interaction between the magnetic
field lines and the fallback disk is ignored.


 According to the law of rotation, the time derivative of the NS spin can be calculated from
\begin{equation}
 \dot{\Omega}=-\frac{2\dot{M}R_{\rm m}^{2}}{I}\left ( \Omega-\Omega_{\rm K}(R_{\rm m}) \right ) =-\frac{\Omega-\Omega_{\rm K}
 (R_{\rm m})}{t_{\rm prop}}, \label{eq:propeller}
\end{equation}
where
\begin{equation}
	t_{\rm prop}=28.2{~\rm yr} I_{45} \xi_{0.5}^{-2} M_{1.4}^{2/7} B_{14}^{-8/7} R_{6}^{-24/7} \dot{M}_{18}^{-3/7}.
\end{equation}
It is worth noting that $t_{\rm prop}$ changes with time in our model, which is different with the model of \cite{ho17}.
Therefore, we have to use a numerical method to calculate the spin evolution of the NS in the propeller phase.

\section{Evolutionary models of ultra-long period radio transients}
\subsection{Initial values and input parameters}
In the ejector phase, we use Equation (6) to calculate the spin evolution of the NS. Once the NS transitions to the
propeller phase, we adopt a numerical method based on Equations (13) and (14) to simulate its spin evolution.
For simplicity, we take $\xi_{0.5}=M_{1.4}=R_{6}=I_{45}=1$. The spin period of the nascent NS is thought to be $P_{0}=10~\rm ms$ ($P_{0,-3}=10$). Adopting the above input parameters, the spin evolution
of the NS is governed by two input parameters including its magnetic field $B$ and the initial mass-accretion rate
$\dot{M}_{\rm 0}$ of the fallback disk.

\subsection{Radio transient J1627}

\begin{figure}
\centering
\includegraphics[width=1.15\linewidth,trim={0 0 0 0},clip]{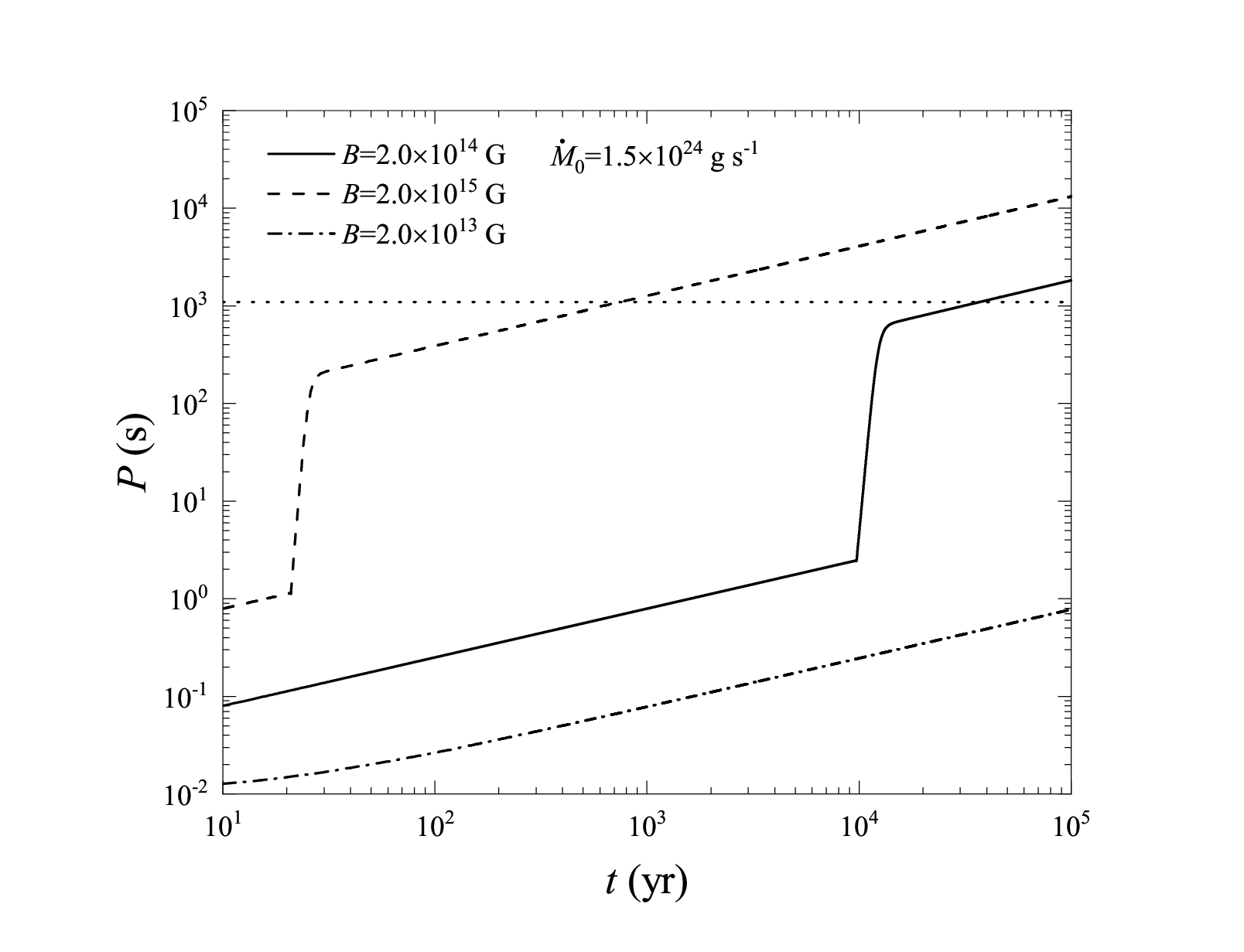}
\caption{Evolution of spin periods of NSs surrounding a fallback disk with an initial mass-accretion rate
$\dot{M}_{0}=1.5\times10^{24}~\rm g\,s^{-1}$ for different surface magnetic fields.
The solid, dashed, and dashed-dotted curves denote the evolutionary tracks of NSs with $B=2.0 \times10^{14}$
, $2.0\times10^{15}$, and $2.0\times10^{13}~\rm G$, respectively. The horizontal dotted line represents the present period $P=1091~\rm s$ of radio
transient J1627.} \label{fig:P1627}
\end{figure}

\begin{figure}
\centering
\includegraphics[width=1.15\linewidth,trim={0 0 0 0},clip]{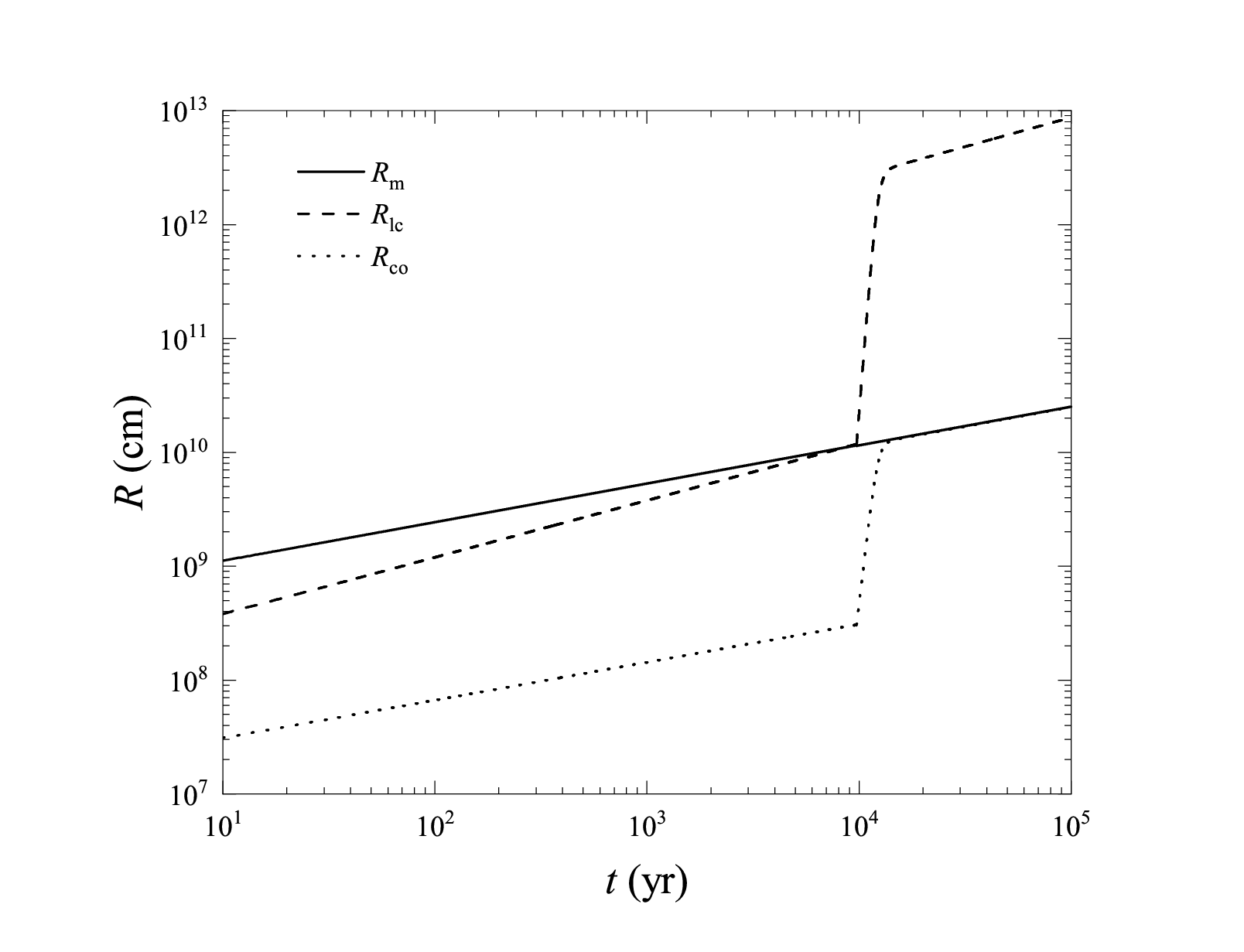}
\caption{Evolution of the three critical radii of an NS with a fallback disk of an initial mass-accretion
rate $\dot{M}_{0}=1.5\times10^{24}~\rm g\,s^{-1}$ and a surface magnetic field $B=2.0 \times10^{14}~\rm G$.
The solid, dashed, and dotted curves denote the evolutionary tracks of the magnetospheric radius, light-cylinder radius, and
co-rotation radius, respectively. } \label{fig:1627radius}
\end{figure}

\begin{figure}
\centering
\includegraphics[width=1.15\linewidth,trim={0 0 0 0},clip]{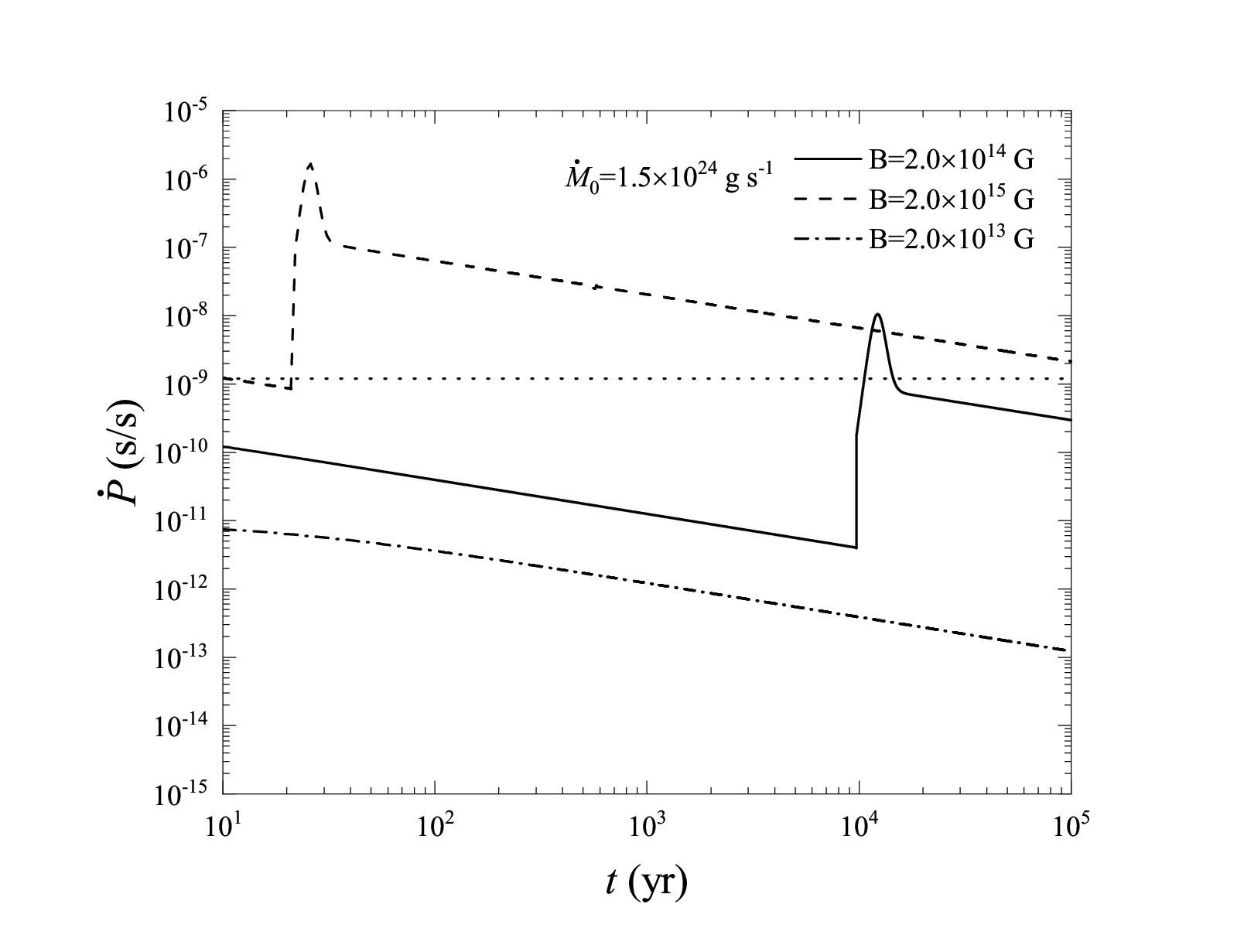}
\caption{Spin-period derivatives of NSs as a function of ages in the existence of a fallback disk with an initial mass-accretion
rate $\dot{M}_{0}=1.5 \times10^{24}\ \rm g\,s^{-1}$. The solid, dashed, and dashed-dotted curves denote the evolutionary tracks of
NSs with $B=2.0 \times10^{14}$, $2.0\times10^{15}$, and $2.0\times10^{13}~\rm G$, respectively. The horizontal dotted line represents
the upper limit ($\dot{P} = 1.2 \times10^{-9}~\rm s\ s^{-1}$) of the period derivative of J1627.} \label{fig:pdot1627}
\end{figure}

Figure \ref{fig:P1627} shows the evolutionary tracks of NSs with an initial mass-accretion rate
$\dot{M}_{0}=1.5\times10^{24}~\rm g\,s^{-1}$ and different surface magnetic fields in the spin period versus
NS age diagram. In the ejector phase, the spin periods evolve along lines with the same slope of
$\bigtriangleup{\rm log}(P/{\rm s})/\bigtriangleup{\rm log}(t/ \rm yr)=1/2$. This law arises from
a simple relation $P\approx P_{0}(t/t_{\rm em})^{1/2}$ when $t/t_{\rm em}\gg1$ according to Equation (\ref{eq:ejectorP}). The
final period in the ejector phase is $P_{\rm ej,max}$, and the duration of the ejector phase $t_{\rm ej}\propto B_{14}^{-8/3}$
for a fixed $\dot{M}_{0}$. Therefore, a strong magnetic field naturally results in a short duration of the ejector phase like in Figure 1. In contrast, the NS with a weak magnetic field of $2.0\times10^{13}$ G is always in the ejector phase in a timescale of $10^{5}$ yr, and merely evolves to a period of 0.8 s. Furthermore, from Equations (2), (8),
and (10), we can derive the evolutionary law of $P_{\rm ej,max}$ as
\begin{equation}
P_{\rm ej,max}\propto B_{14}^{4/7}t_{\rm ej}^{19/56}\propto B_{14}^{-1/3}.
\end{equation}
As a consequence, a strong magnetic field produces a short $P_{\rm ej,max}$. After $P_{\rm ej,max}$, the NS transitions to the
propeller phase, and its spin period increases at an exponential rate. The two NSs with strong magnetic field
$B=2.0\times10^{15}$ and $2.0\times10^{14}$ G can evolve to the present period (1091 s) of J1627 at
$t=748$ and $3.7\times10^{4}$ yr, respectively.

The model with an initial mass-accretion rate $\dot{M}_{0}=1.5\times10^{24}~\rm g\,s^{-1}$ and a surface magnetic field
$B=2.0 \times10^{14}~\rm G$ can successfully reproduce the observed $P$ and $\dot{P}$ of J1627.
We depict the evolution of the three critical radii of the NS in Figure \ref{fig:1627radius}. During the ejector phase,
$R_{\rm m}>R_{\rm lc}>R_{\rm co}$, and the NS spins down due to magnetic dipole radiation. At $t=9.7\times10^{3}~\rm yr$,
$R_{\rm m}=R_{\rm lc}>R_{\rm co}$, and the NS transitions to the propeller phase. Subsequently, $R_{\rm lc}>R_{\rm m}>R_{\rm co}$,
and the spin period of the NS rapidly increases at an exponential rate due to the propeller torque. With the increase of the spin
period, the corotation radius also increases at an exponential rate. When $R_{\rm co}$ increases to
be approximately equal to $R_{\rm m}$, the NS reaches a quasi-spin equilibrium, and the quasi-equilibrium
period is given by
\begin{equation}
P_{\rm eq}=9.3~{\rm s}~ B_{14}^{6/7}\dot{M}_{18}^{-3/7}\propto t^{57/112}.
\end{equation}
As a consequence, the spin period of the NS slowly increases in the quasi-spin-equilibrium stage. Because
the slope of the evolutionary track is $\bigtriangleup{\rm log}(P_{\rm eq}/{\rm s})/\bigtriangleup{\rm log}(t/ \rm yr)=57/112\approx1/2$
, it seems that the two evolutionary lines of spin periods in the ejector phase and the
quasi-spin-equilibrium stage are parallel (see also Figure 1).

Figure \ref{fig:pdot1627} plots the evolutionary tracks of NSs with $\dot{M}_{0}=1.5\times10^{24}~\rm g\,s^{-1}$ and
different surface magnetic fields in the spin-period derivative versus NS age diagram. It is clear that the evolutionary tracks in the ejector
phase are lines with the same slope. This phenomenon is caused by the evolutionary law of spin-period derivative. In the ejector phase, the evolution of the period derivative is governed by
\begin{equation}
\dot{P}=\frac{4\pi^{2}\beta}{P}=\frac{4\pi^{2}\beta}{P_{0}}(1+\frac{t}{t_{\rm em}})^{-1/2}.\label{eq:ejectorpdot}
\end{equation}
When $t/t_{\rm em}\gg1$, Equation (\ref{eq:ejectorpdot}) can derive to $\dot{P}\propto(t/t_{\rm em})^{-1/2}$, hence the spin-period derivatives evolve along lines with a slope of $\bigtriangleup{\rm log}(\dot{P}/{\rm s\,s^{-1}})/\bigtriangleup{\rm log}(t/ \rm yr)=-1/2$.
The period derivatives decrease to $\dot{P}\sim 10^{-12}-10^{-9}~\rm s\,s^{-1}$, depending on the surface magnetic fields of NSs. A strong surface magnetic
field tends to result in a high final period derivative in the ejector phase.

Once the NS transitions to the propeller phase, the spin-period derivatives rapidly climb to a peak, and then decline. The evolutionary law of the period derivative is
\begin{equation}
\dot{P}=-\frac{2\pi\dot{\Omega}}{\Omega^{2}},
\end{equation}
where $\dot{\Omega}$ is derived from Equation (\ref{eq:propeller}). Therefore, a rapidly decreasing angular velocity produces a quick
increasing $\dot{P}$. For the NS with $B=2.0\times10^{14}~\rm G$, our simulated $\dot{P}=4.8\times10^{-10}~\rm s\,s^{-1}$ at the
current age ($3.7\times10^{4}~\rm yr$) of J1627, which is less than the observed upper limit ($1.2\times10^{-9}~\rm s\,s^{-1}$) of period derivative of J1627 \citep{hurl22}. Based on the observed period and period derivative, the characteristic age of J1627 can be constrained to be $\tau_{\rm c}=P/(2\dot{P})>1.4\times10^{4}~\rm yr$. Our simulated age is compatible with the characteristic age of J1627. The NS with a strong magnetic field of $B=2.0\times10^{15}~\rm G$ can evolve to 1091 s at an age of 748 yr, while the corresponding period derivative is much higher than the observed value. From Equation (16), it can derive $\dot{P}_{\rm eq}\propto t^{-55/112}$, thus, the slope of the evolutionary lines of $\dot{P}_{\rm eq}$ is $\bigtriangleup{\rm log}(\dot{P}_{\rm eq})/\bigtriangleup{\rm log}(t/ \rm yr)=-55/112\approx-1/2$. As a consequence, it seems
that the two evolutionary lines of period derivatives in the ejector phase and the quasi
-spin-equilibrium stage are parallel in Figure 3.

In our best model, the current mass-accretion rate of J1627 is $\dot{M}=\dot{M}_{0}(t/2000~\rm s)^{-19/16}=5.8\times10^{13}~\rm g\,s^{-1}$ when its
current age $t=3.7\times10^{4}~\rm yr$. Therefore, the current X-ray luminosity of the NS can be calculated by
\begin{equation}
\begin{aligned}
L_{\rm X}=\frac{GM\delta\dot{M}_{\rm c}}{R}=1.0\times10^{32}~{\rm erg\,s^{-1}}M_{1.4}R_{6}^{-1}\\
\left(\frac{\delta}{0.01}\right)
\left(\frac{\dot{M}_{\rm c}}{5.8\times10^{13}~\rm g\,s^{-1}}\right),
\end{aligned}
\end{equation}
where $\delta$ is the accretion efficiency of the NS in the propeller phase. Since J1627 was observed an upper limit
($L_{\rm X}<1.0\times10^{32}~{\rm erg\,s^{-1}}$) of X-ray luminosity \citep{hurl22}, it implies that the accretion
efficiency of the NS is smaller than $1\%$ if the observed X-ray luminosity originates from an accretion from the fallback disk. Such an accretion efficiency is slightly lower than the estimated values ($0.01-0.05$)
in the observation and theoretical researches \citep{cui97,zhan98,papi15,tsyg16}.

\begin{figure}
\centering
\includegraphics[width=1.15\linewidth,trim={0 0 0 0},clip]{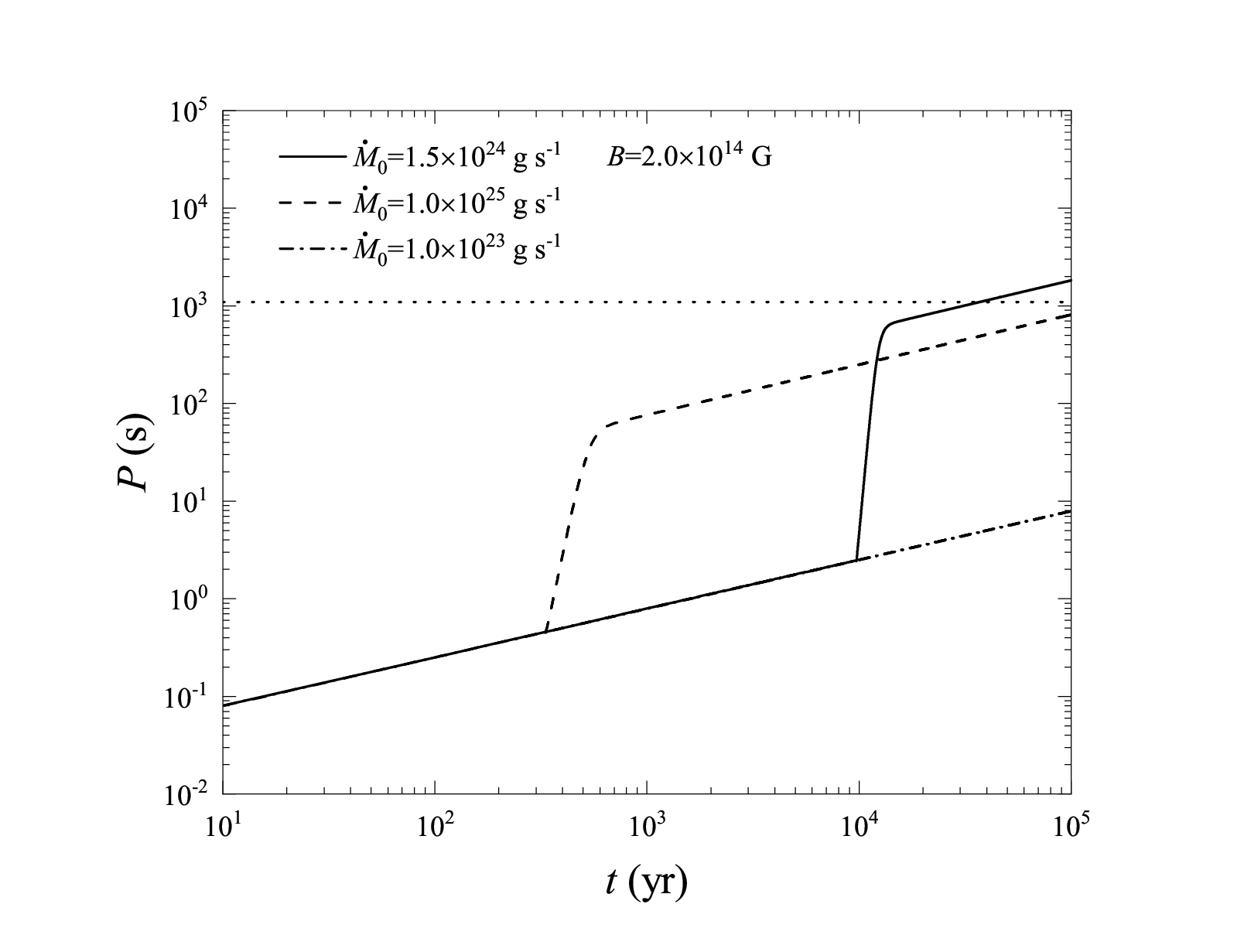}
\caption{Evolution of spin periods of NSs surrounding a fallback disk with different initial mass-accretion rates and
a surface magnetic field $B=2.0 \times10^{14}$ G.
The solid, dashed, and dashed-dotted curves denote the initial mass-accretion rates of the fallback disks $\dot{M}_{0}=1.5 \times10^{24}$, $1.0\times10^{25}$, and $1.0\times10^{23}~\rm g\,s^{-1}$, respectively. The horizontal dotted line represents the present period $P=1091~\rm s$ of radio
transient J1627.} \label{fig:orbmass}
\end{figure}
Figure 4 presents the influence of the initial mass-accretion rates of fallback disks on the spin evolution of NSs. Because of
the same magnetic field, the spin periods of NSs with different initial mass-accretion rates of fallback disks evolve along the same
line in the ejector phase, while their durations are different. From Equation (10), it can derive
$t_{\rm ej}\propto \dot{M}_{0}^{-16/9}$, thus, a high initial mass-accretion rate results in a short duration in the ejector phase.
After the NSs transition to the propeller phase, they quickly reach a quasi-spin equilibrium. A high initial mass-accretion rate
naturally produces a short initial quasi-equilibrium period according to Equation (16). Therefore, the NS with $\dot{M}_{0}=1.0\times10^{25}~\rm g\,s^{-1}$ spends a timescale longer than the NS with $\dot{M}_{0}=1.5\times10^{24}~\rm g\,s^{-1}$ to evolve to the current period of J1627.

\begin{figure}
\centering
\includegraphics[width=1.15\linewidth,trim={0 0 0 0},clip]{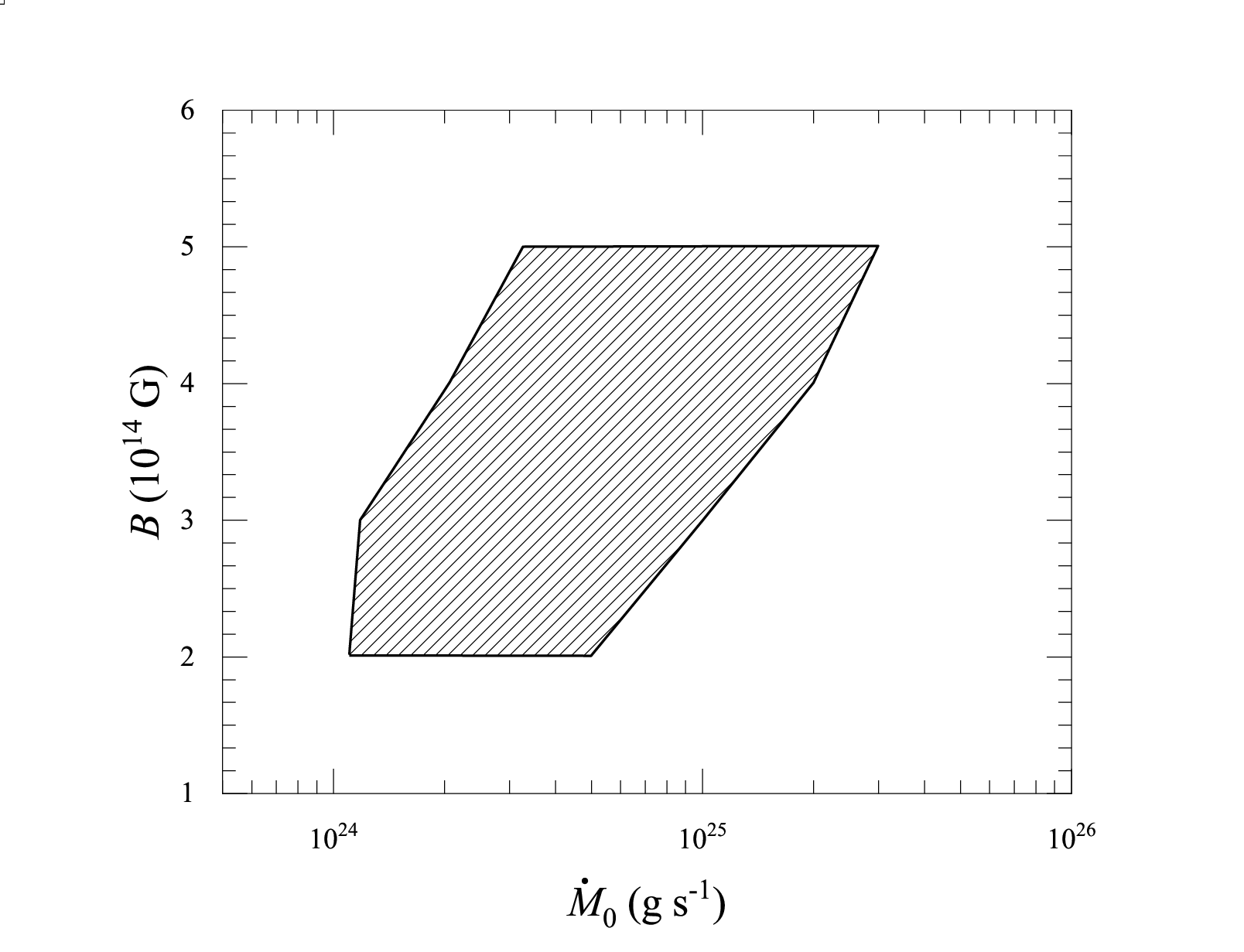}
\caption{Parameter space that can produce the ultra-long period radio transient J1627 in magnetic field vs. initial mass-accretion
rate diagram. Those nascent NSs with a magnetic field and an initial mass-accretion rate
in the shaded region can evolve into the radio transient J1627 with a spin period of
$1091~\rm s$, a period derivative less than $ 1.2 \times10^{-9}~\rm s\,s^{-1}$, and an X-ray luminosity less than $ 10^{32}~\rm erg\,s^{-1}$ in a timescale shorter than $10^{5}~\rm yr$.} \label{fig:parameter}
\end{figure}

Figure \ref{fig:parameter} summarizes the parameter space that can produce the ultra-long period radio transient J1627
in the magnetic field versus the initial mass-accretion rate diagram. Those magnetars with a magnetic field of $(2-5)\times 10^{14}~\rm G$ and a fallback disk with an initial mass-accretion rate of $(1.1-30)\times10^{24}~\rm g\,s^{-1}$ can evolve toward the radio transient J1627 with a spin period of $1091~\rm s$, a period derivative less than $ 1.2 \times10^{-9}~\rm s\,s^{-1}$, and an X-ray luminosity less than $ 10^{32}~\rm erg\,s^{-1}$ (taking a relatively low accretion efficiency $\delta=0.001$, we calculate the X-ray luminosity of the NS in the propeller phase according to equation 19) in a timescale shorter than $10^{5}~\rm yr$. It is worth emphasizing that the parameter space strongly depends on the accretion efficiency during the propeller phase. The parameter space would sharply reduce to an ultra-small zone in the lower-left shaded area if the accretion efficiency $\delta=0.01$. To form the ultra-long spin period, a strong magnetic field tends to require a high initial mass-accretion rate. If $t_{\rm prop}$ is a constant, it can derive
to $\Omega=[\Omega_{\rm ej,max}-\Omega_{\rm k}(R_{\rm m})]e^{-(t-t_{\rm ej})/t_{\rm prop}}+\Omega_{\rm k}(R_{\rm m})$
from Equation (\ref{eq:propeller}). Therefore, $t_{\rm prop}$ is a characteristic timescale that the spin period
changes in the propeller phase. A suitable $t_{\rm prop}$ will determine whether the NS can evolve to an ultra-long
spin period. According to Equation (14), a strong magnetic field naturally requires a high initial mass-accretion rate for a
fixed $t_{\rm prop}$. According to Figure 4, a high initial mass-accretion rate can not produce ultra-long period NS in a timescale shorter than $10^{5}~\rm yr$, resulting in the right boundary of the shaded region. An NS with a strong magnetic field would evolve toward the state exceeding the upper limits of the observed period derivative and X-ray luminosity, resulting in the upper boundary. Meanwhile, the bottom and left boundaries originate from the lower limits of the magnetic field and initial mass-accretion rate, under which the NS is always in the ejector phase.

\subsection{Radio transient J1839}
\subsubsection{J1839 is in the propeller phase}
We first consider that J1839 is still in the propeller phase at present. Figure 6 shows the evolutionary trajectories of $P$ and $ \dot{P}$ of NSs with the initial mass-accretion rate
$\dot{M}_{0}=1.0\times10^{25}~\rm g\,s^{-1}$ and different magnetic fields. The NS with a weak magnetic field of
$B=1.0\times 10^{12}~\rm G$ is permanently in the ejector phase in a timescale of $10^{8}~\rm yr$, and cannot evolve to an ultra-long period. The two NSs with magnetic fields of $B=1.0\times 10^{14}$ and $7.9\times 10^{12}$ G can evolve to the current period of J1839 in ages of $8.3\times10^{5}$ and $6.0\times10^{7}$ yr, respectively. In the case of $B = 7.9\times 10^{12}~\rm G$, the current period derivative $\dot{P}=3.55\times10^{-13}~\rm s\,s^{-1}$. However, the NS with $B = 1.0\times 10^{14}~\rm G$ spends a timescale longer than its current age to evolve to the observed upper limit ($\dot{P}=3.6\times10^{-13}~\rm s\,s^{-1}$) of the period derivative.

Similar to J1627, we also investigate whether it can produce the observed properties of J1839 in a wide range of magnetic fields and initial mass-accretion rates. Those NSs with initial mass-accretion rates of $\dot{M}_{0}=10^{23}-10^{30}~\rm g\,s^{-1}$ and magnetic fields of $B=10^{9}-10^{16}~\rm G$ cannot evolve to the current period and period derivative ($\dot{P}<3.6\times10^{-13}~\rm s\,s^{-1}$) in a timescale shorter than $10^{7}~\rm yr$. It is generally thought that the active lifetime of a fallback disk is $\sim 10^{5}$ yr \citep{genc22}, thus, the probability that J1839 is in the propeller phase can be ruled out.

\subsubsection{J1839 is in the second ejector phase}
J1839 may experience the first ejector phase, and the propeller phase in a timescale of $10^{5}~\rm yr$ that the fallback disk is active. Subsequently, the source transitions to the second ejector phase again because the fallback disk becomes inactive. Figure 7 shows the evolutionary tracks of $P$ and $ \dot{P}$ of NSs with an initial mass-accretion rate
$\dot{M}_{0}=4.0\times10^{24}~\rm g\,s^{-1}$ and different magnetic fields. The NS with a weak magnetic field of $B=1.0\times10^{13}~\rm G$ is consistently in the ejector phase and evolves to a period of $\sim 10$ s in a timescale of $10^{8}$ yr. The NS with a strong magnetic field of $B=1.0\times10^{15}~\rm G$ can evolve to the present period of J1839 in the quasi-spin equilibrium stage, while the corresponding period derivative is much higher than the observed value.

Only the NS with $B=2.0\times10^{14}~\rm G$ successively experiences the first ejector phase, the propeller phase, the quasi-spin equilibrium stage, and the second ejector phase. When the NS age of $t=10^{5}~\rm yr$, the NS was spun down to a period of 1200 s in the quasi-spin equilibrium stage. Subsequently, it transitions to the second ejector phase because the fallback disk becomes inactive. Because the initial period $P_{0}=1200~\rm s$ in the second ejector phase, the $t_{\rm em}$ is approximately 10 orders of magnitude longer than that in the first phase. Since $t\ll t_{\rm em}$, the period of the NS increases at an extremely small rate, which is consistent with the low $\dot{P}$ observed in J1839. At $t=2.5\times10^{7}~\rm yr$, the period of the NS increases to the current period (1318 s) of J1839, and the period derivative decreases to be $\dot{P}=8.0\times10^{-14}~\rm s\,s^{-1}$, which is less than the observed upper limit of $\dot{P}=3.6\times10^{-13}~\rm s\,s^{-1}$. At $t=10^{5}~\rm yr$, the propeller torque exerted on the NS changes into the magnetic dipole radiation torque, thus, the $\dot{P}$ sharply decreases.

\begin{figure}
\centering
\includegraphics[width=1.15\linewidth,trim={0 0 0 0},clip]{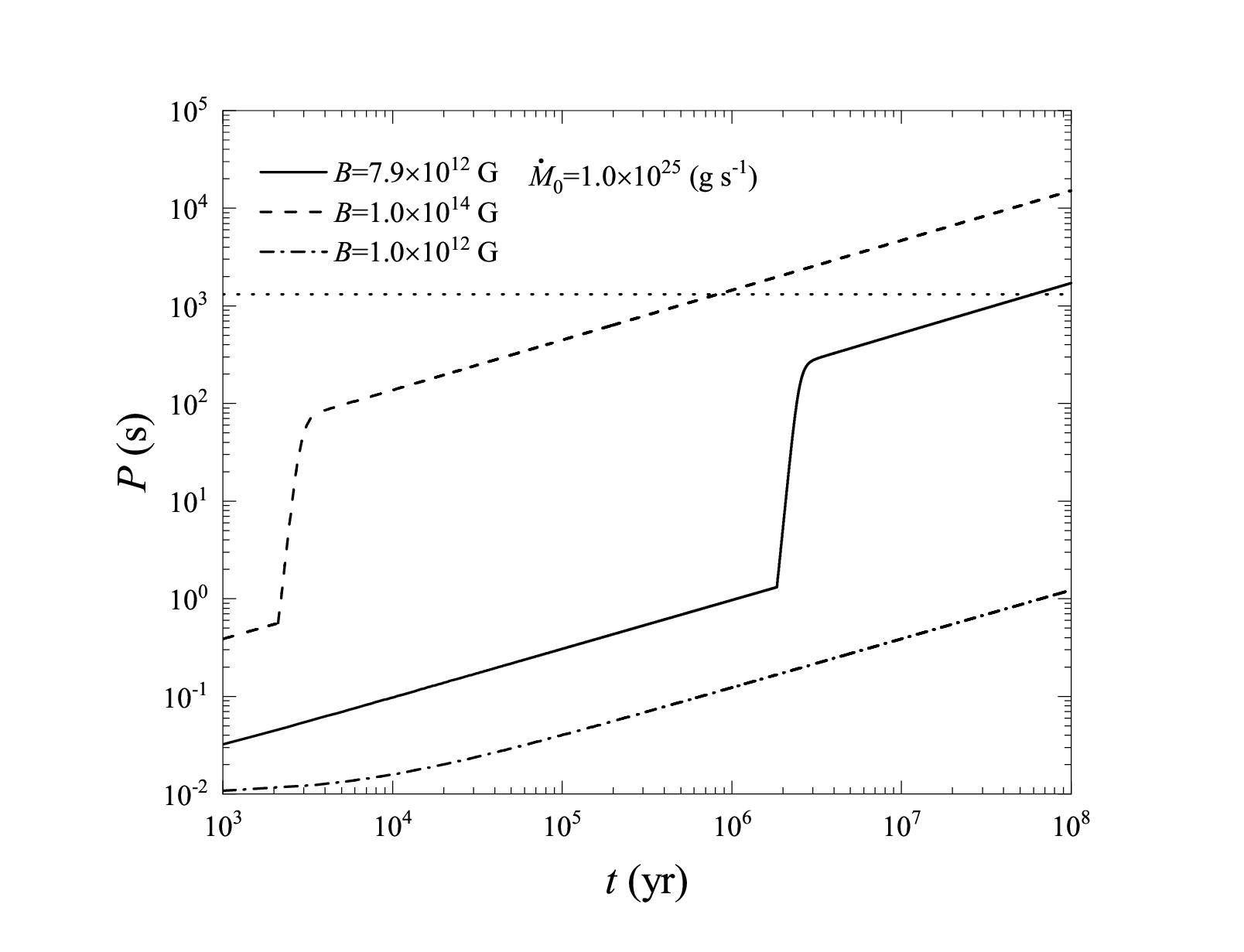}
\includegraphics[width=1.15\linewidth,trim={0 0 0 0},clip]{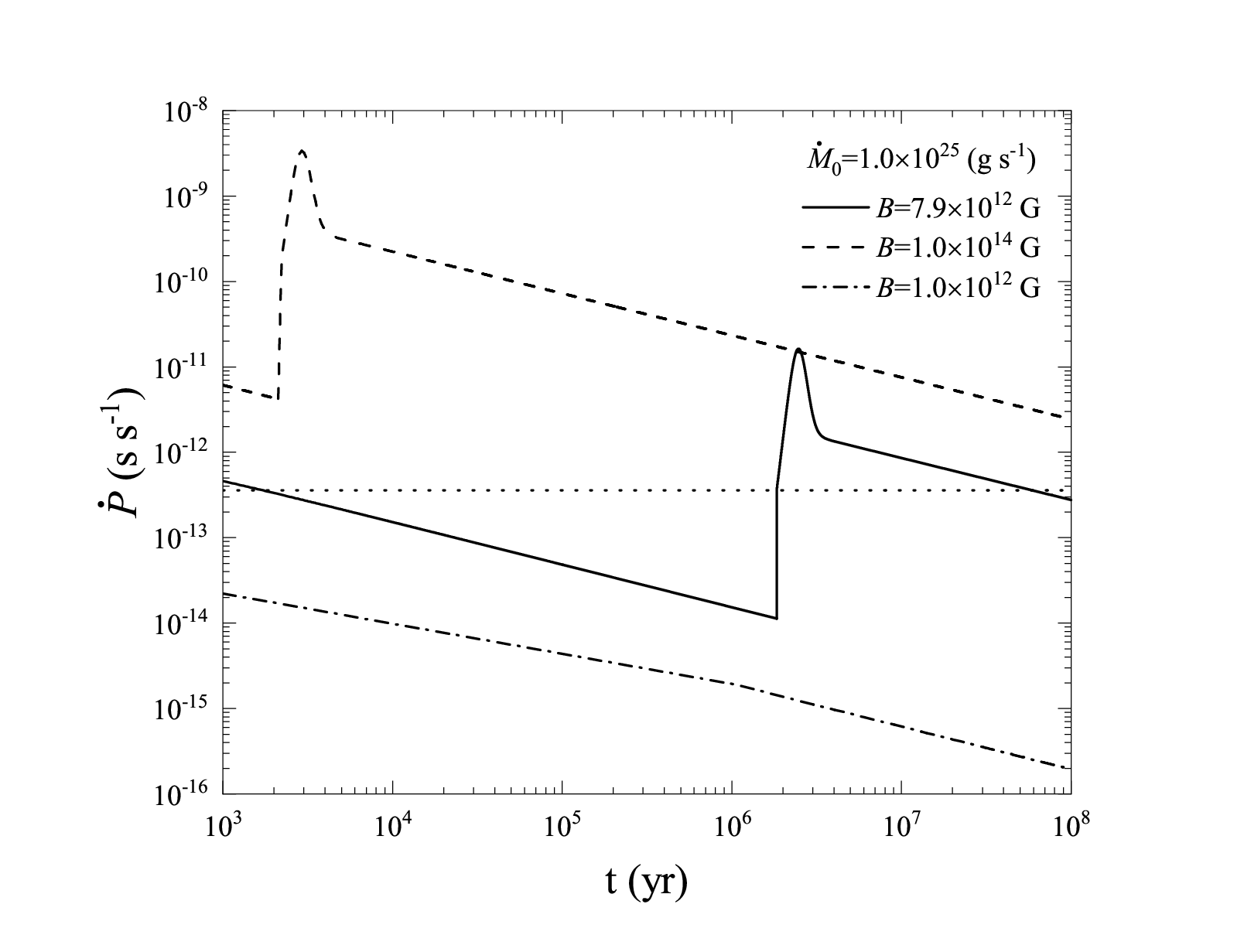}
\caption{Evolution of NSs with a fallback disk of an initial mass-accretion rate
$\dot{M}_{0}=1.0\times10^{25}~\rm g\,s^{-1}$ and different surface magnetic fields in the spin period vs. NS age
diagram (top panel) and period derivative vs. NS age diagram (bottom panel). The fallback disks are assumed to be always active. The horizontal dotted line in the top panel and bottom panel represents the present period $P=1318~\rm s$ and the upper limit
($\dot{P} = 3.6 \times10^{-13}~\rm s\ s^{-1}$) of the period derivative of radio transient J1839, respectively. } \label{fig:orbmass}
\end{figure}

\begin{figure}
\centering
\includegraphics[width=1.15\linewidth,trim={0 0 0 0},clip]{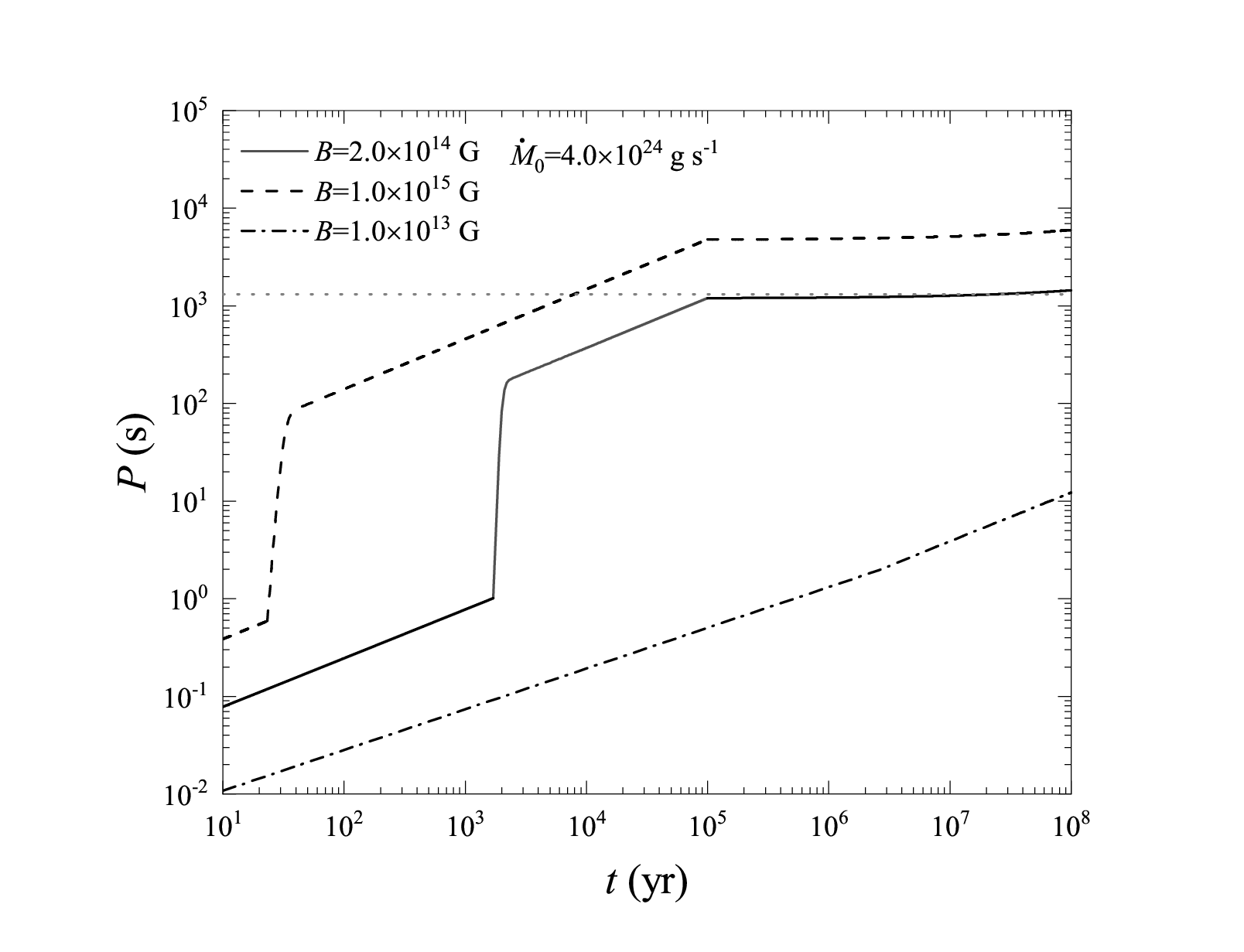}
\includegraphics[width=1.15\linewidth,trim={0 0 0 0},clip]{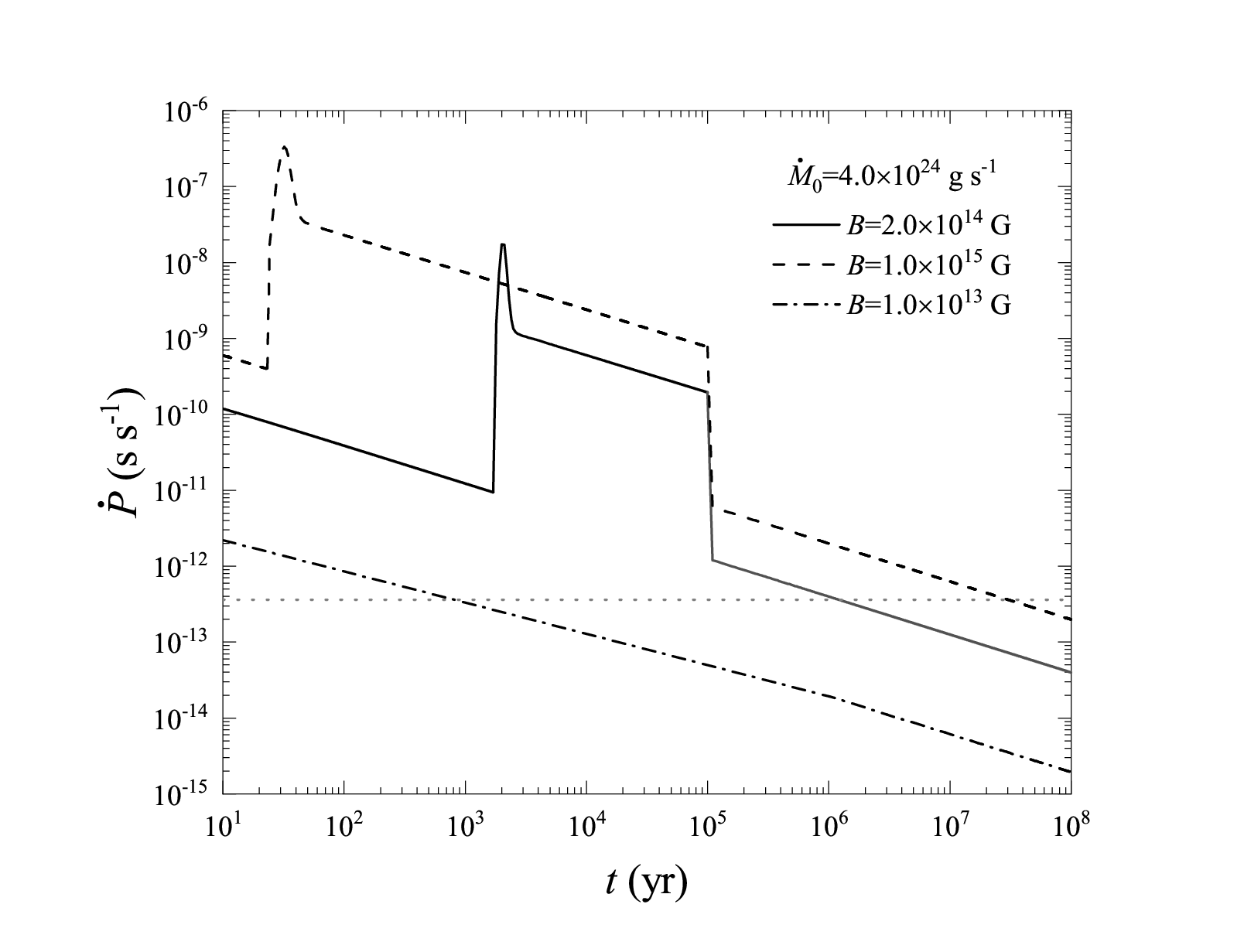}
\caption{Same as Figure 6 but for $ \dot{M}_{0}=4.0\times10^{24}~\rm g\,s^{-1}$ and $B=1.0\times10^{13}, 2.0\times10^{14}$,
and $1.0\times10^{15}$ G. Furthermore, the fallback disks are thought to be inactive after the NS age is greater than $10^{5}~\rm yr$.} \label{fig:orbmass}
\end{figure}

To understand the influence of $\dot{M}_{0}$ on the spin evolution of NSs, we also chart the evolution of the spin periods of NSs with $B=2.0\times10^{14}$ G and $\dot{M}_{0}=4.0\times10^{23}, 4.0\times10^{24}$, and $4.0\times10^{25}~\rm g\,s^{-1}$ in Figure 8.
Since the duration of the first ejector phase $t_{\rm ej}\propto \dot{M}_{0}^{-16/9}$, the NS with a small initial mass-accretion rate of $4.0\times10^{23}~\rm g\,s^{-1}$ cannot transition to the propeller phase in a timescale of $10^{8}~\rm yr$. However, the NS with a high initial mass-accretion rate of $4.0\times10^{25}~\rm g\,s^{-1}$ enters the propeller phase in a short timescale of $\sim 30~\rm yr$.

\begin{figure}
\centering
\includegraphics[width=1.15\linewidth,trim={0 0 0 0},clip]{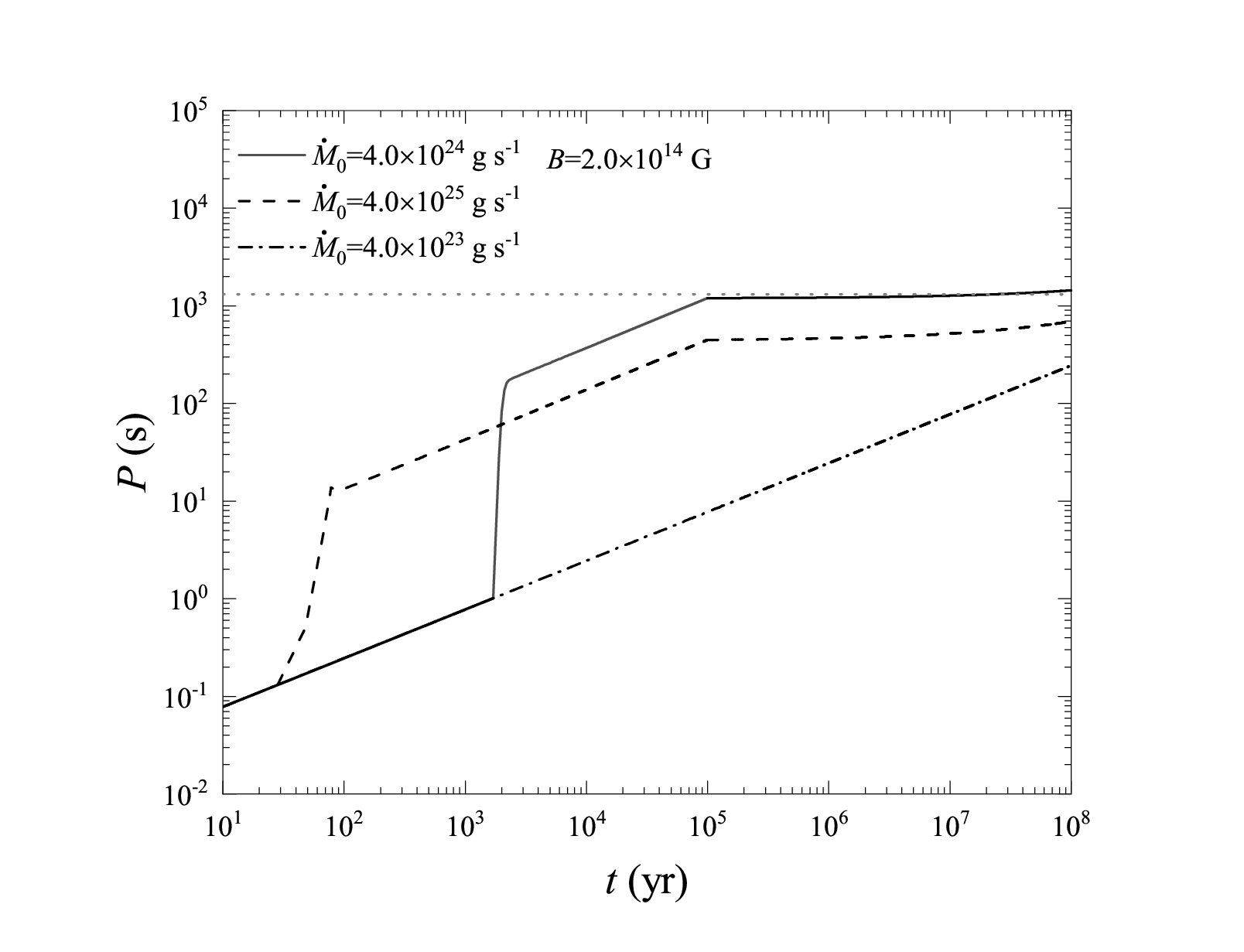}
\caption{Same as the top panel of Figure 7 but for $B=2.0\times10^{14}$ G and $\dot{M}_{0}=4.0\times10^{23}, 4.0\times10^{24}$, and $4.0\times10^{25}~\rm g\,s^{-1}$.} \label{fig:orbmass}
\end{figure}

To investigate the progenitor properties of J1839, in Figure 9 we summarize the parameter space that can produce the ultra-long period radio transient J1839 in magnetic field versus the initial mass-accretion rate diagram. The characteristic age of J1839 is $\tau_{\rm c}=P/(2\dot{P})>1.2\times10^{8}~\rm yr$. In the simulation, we diagnose whether an NS can evolve to the current state of J1839 in a timescale of $1.2\times10^{8}~\rm yr$. It is clear that the potential progenitor of J1839 is also a magnetar. Those magnetars with a magnetic field of $(2-6)\times 10^{14}~\rm G$ and a fallback disk with an initial mass-accretion rate of $4.0\times10^{24}-10^{26}~\rm g\,s^{-1}$ are possible progenitors of radio transient J1627. Similar to J1627, a strong magnetic field tends to require a high initial mass-accretion rate in order to evolve into J1839. Furthermore, an NS with a magnetic field stronger than $6.0\times10^{14}~\rm G$ can evolve to the observed period of J1839, while the corresponding period derivative is much higher than the observed value.

\section{Discussion}
\begin{figure}
\centering
\includegraphics[width=1.15\linewidth,trim={0 0 0 0},clip]{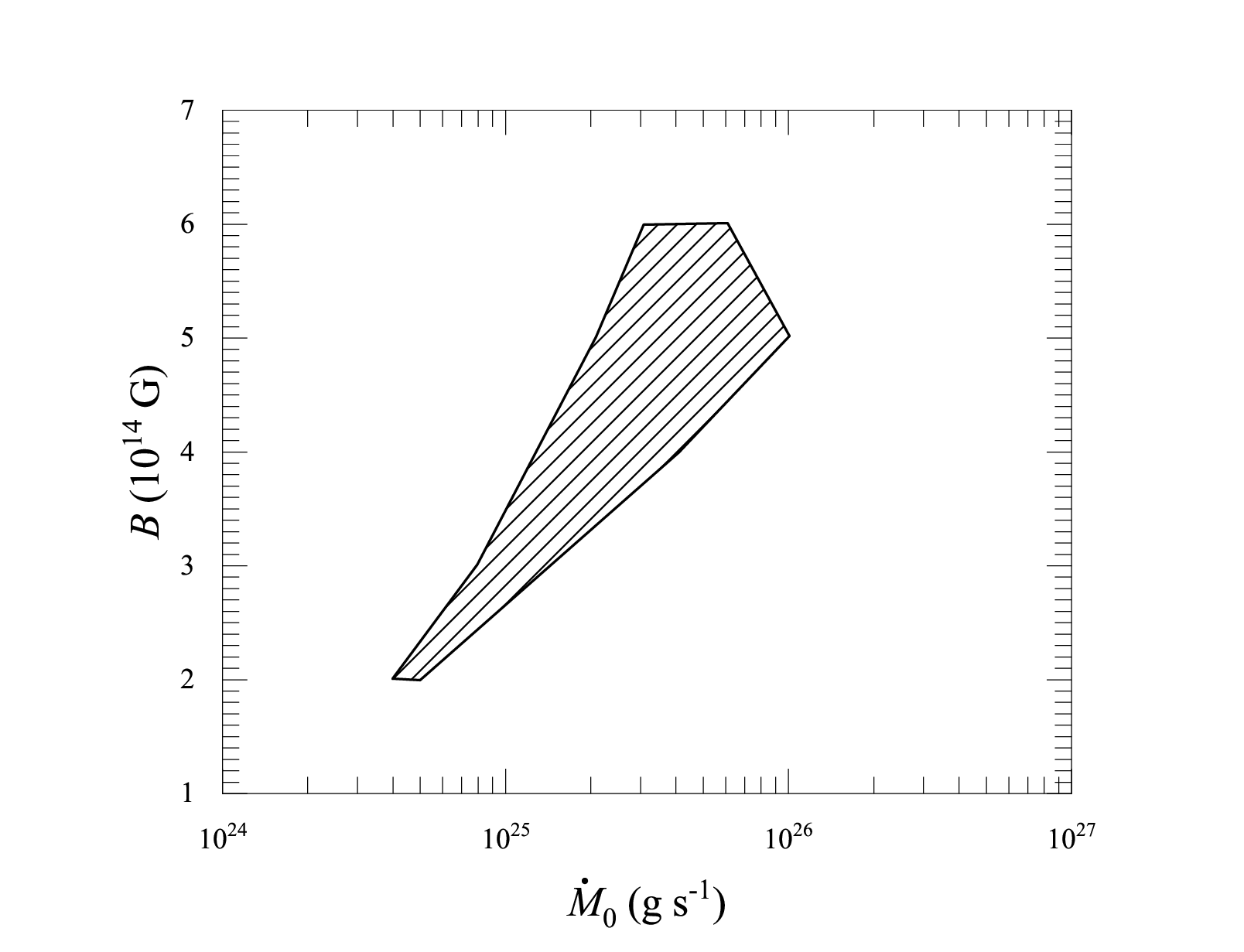}
\caption{Parameter space that can produce the ultra-long period radio transient J1839 in magnetic field vs. initial mass-accretion
rate diagram. Those nascent NSs with a magnetic field and an initial mass-accretion rate
in the shaded region can evolve into the radio transient J1839 with a spin period of
$1318~\rm s$ and a period derivative less than $ 3.6\times10^{-13}~\rm s\,s^{-1}$ in a timescale shorter than $1.2\times10^{8}~\rm yr$. The fallback disks are assumed to be inactive after the NS age is greater than $10^{5}~\rm yr$.} \label{fig:orbmass}
\end{figure}
\subsection{Comparison with previous works}
Assuming an initial spin period $P_{0} = 10~\rm ms$ and an initial mass-accretion rate
$\dot{M}_{0} \sim 10^{23}~\rm  g\,s^{-1}$, \cite{ronc22} found that a magnetar with initial magnetic field $B_{0} \sim10^{14}$ G
can spin down to a spin period of 1091 s in a timescale of $10^{3}-10^{5}$ yr. However, our models require a relatively high
initial mass-accretion rate of $\ga 10^{24}~\rm  g\,s^{-1}$. This discrepancy should arise from their different accretion model of
the fallback disk, in which $\dot{M}=\dot{M}_{0}(1+t/T)^{-\alpha}$ \citep{meno01,erta09}. Furthermore, they considered the decay of
magnetic fields under the combined contribution of ohmic dissipation and the Hall effect in the NS crust, which causes
the NS to slightly spin up after the spin equilibrium. However, our simulated spin period slowly increases during
the quasi-spin equilibrium.

Adopting a same model in \cite{tong16}, \cite{tong23a} found that a magnetar with a magnetic field
$B=4.0\times10^{14}$ G can be spun down to 1091s by a self-similar fallback disk with initial mass of
$M_{\rm d,i}=10^{-3}-10^{-4}~M_{\odot}$. From $\dot{M}_{0}=(\alpha-1)M_{\rm d,i}/(\alpha T)$, the corresponding initial
mass-accretion rates are $1.6\times(10^{25}-10^{26})~\rm g\,s^{-1}$. According to our simulated parameter space forming J1627,
an NS with $\dot{M}_{0}=1.6\times10^{25}~\rm g\,s^{-1}$ and $B=4.0\times10^{14}$ G can evolve into J1627, while an NS
with $\dot{M}_{0}=1.6\times10^{26}~\rm g\,s^{-1}$ and $B=4.0\times10^{14}$ G is unlikely to evolve toward J1627 in a
timescale less than $10^{5}~\rm yr$. This discrepancy may be caused by different determining criterion forming J1627, in which
the spin period is a unique criterion in \cite{tong23a}, while both $P$ and $\dot{P}$ are criterions forming J1627 in our simulations.

\cite{genc22} shown that an NS with an initial period of 0.3 s, a magnetic field of $\sim10^{12}~\rm G$, and a fallback disk with an initial mass of $1.6\times10^{-5}~M_{\odot}$ can evolve into J1627. Their simulations can interpret the observed period, period derivative, and X-ray luminosity of J1627 when the disk becomes completely inactive in a timescale of $7\times10^{5}~\rm yr$. Our magnetic field is much stronger than that in their model, and the evolutionary timescale is one order of magnitude smaller than their result. Different torque models should be responsible for these discrepancies.

\subsection{Evolutionary fates of NSs under different input parameters}
According to the NS $+$ fallback disk model, the evolutionary fates of pulsars depend on the two input parameters: magnetic field $B$ and initial mass-accretion rate $\dot{M}_{0}$ of the fallback disk. In the $\dot{P}-P$ diagram of pulsars, we depict the evolutionary tracks of NSs with several special $B$ and $\dot{M}_{0}$. An NS with a weak magnetic field of $2.0\times10^{13}~\rm G$ is always in the ejector phase and evolves into a normal pulsar. The two NSs with $B=2.0\times 10^{14}~\rm G$, and $\dot{M}_{0}=1.5\times10^{24}$ and $4.0\times10^{24}~\rm g\,s^{-1}$ evolve toward radio transients with an ultra-long period and a relatively high $\dot{P}\sim 10^{-10}-10^{-9}\rm s\,s^{-1}$ in the propeller phase. Such a $\dot{P}$ is compatible with the observed upper limit of J1627, hence J1627 is probably in the propeller phase. However, J1839 has a very small period derivative as $\dot{P}<3.6\times10^{-13}~\rm s\,s^{-1}$. Therefore, the evolutionary state of J1839 should be different with J1627. Assuming a fallback-disk active timescale of $10^{5}~\rm yr$, our models show that the magnetic dipole torque in the second ejector phase can produce a $\dot{P}$ that is compatible with the observed upper limit of J1839.

J0901 is a long period pulsar with a period of $P = 75.9$ s and a period derivative of $\dot{P}=2.25\times10^{-13}~\rm s\,s^{-1}$ \citep{cale22}. If this NS is spinning down by a pure magnetic dipolar radiation, the dipolar magnetic field can be estimated to be $B=1.3\times 10^{14}~\rm G$ \citep{ronc22}. Our simulations indicate that an NS with a strong magnetic field of $B=1.0\times 10^{14}~\rm G$ and a high initial mass-accretion rate of $\dot{M}_{0}=1.0\times10^{27}~\rm g\,s^{-1}$ can evolve to the current state of J0901 in the second ejector phase. Recently, \cite{genc23} found that an NS with a weak magnetic field of $\sim10^{12}~\rm G$ can evolve to the current state of J0901 in the strong propeller phase.

J0250 is another slow-spinning radio pulsar with a spin period of 23.5 s and a spin-period derivative of $\dot{P}=2.7\times10^{-14}~\rm s\,s^{-1}$ \citep{tan18}. In Figure 10, the observed $P$ and $\dot{P}$ of J0250
can be matched by the blue evolutionary track, in which the nascent NS has a magnetic field of $B=4.0\times 10^{13}~\rm G$
and $\dot{M}_{0}=2.0\times10^{28}~\rm g\,s^{-1}$. The required magnetic field in our model is approximately equal to the inferred
one \citep[$B=2.6\times 10^{13}~\rm G$,][]{tan18}. To evolve into the current states of J0901 and J0250, it requires extremely
high initial mass-accretion rates of $10^{27}-10^{28}~\rm g\,s^{-1}$. According to Equation (9), the duration of the first ejector phase $t_{\rm ej}\propto \dot{M}^{-4/7}$, thus, these two sources merely experience a very short timescale in the first ejector.

\begin{figure}
\centering
\includegraphics[width=1.15\linewidth,trim={0 0 0 0},clip]{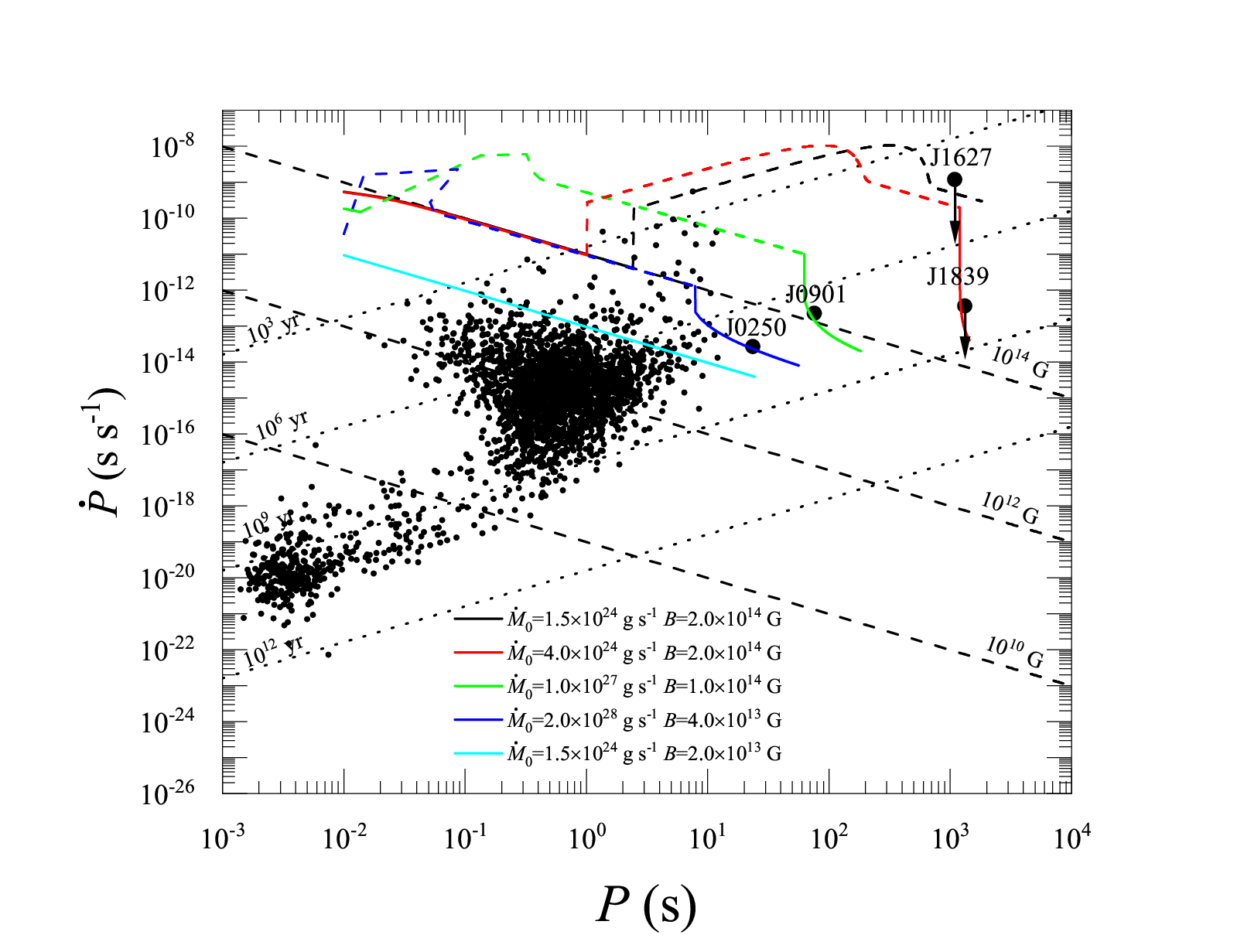}
\caption{Evolutionary fates of NSs with different $B$ and $\dot{M}_{0}$ in the $\dot{P}-P$ diagram. The solid and dashed curves represent the evolutionary tracks in the ejector phase and the propeller phase, respectively. The arrows of J1627 and J1839 stand for their possible range of $\dot{P}$. The fallback disks are assumed to be inactive after the NSs ages are greater than $10^{5}~\rm yr$.} \label{fig:orbmass}
\end{figure}

\subsection{Influence of magnetic field decay}
In the calculations, we ignore a decay of the magnetic field. Actually, the magnetic fields of NSs ought to decay due to the
ohmic dissipation and the Hall effect in their crusts \citep{pons07,pons09,viga13,pons19,gran20}. An evolution simulation shown that the decay of an initial magnetic field of $10^{14}$ G cannot influence the spin-period evolution of the ejector phase in a timescale of $10^{4}$ yr \citep[see also Figure 1 of][]{ronc22}. However, the decay of magnetic fields would produce a long $t_{\rm prop}$ according to Equation (14). As a consequence, J1627 would take a relatively long timescale to evolve to the current period.

If J1839 can evolve to an ultra-long period that is very close to the current period in the active stage of the fallback disk, it would naturally transition to the second ejector phase. Since $t_{\rm em}\propto B^{-2}$ according to Equation (7), a decaying magnetic field leads to a relatively long $t_{\rm em}$. Therefore, the increase of the period is slower than the case without magnetic field decay. Accordingly, the evolutionary timescale of J1839 is also slightly longer than our calculation.

\section{Conclusions}
Recently, a radio transient J1627 with an unusually ultra-long period of 1091 s was reported, and it was proposed to be
an ultra-long-period magnetar \citep{hurl22}. In this work, we attempt to diagnose whether a magnetar with a fallback disk
could be spun down to the current spin period of J1627 by the propeller torque. Our simulations indicate that an NS with a magnetic field of $B=(2-5)\times10^{14}~\rm G$ and a fallback disk with an initial mass-accretion rate of
$\dot{M}_{0}=(1.1-30)\times10^{24}~\rm g\,s^{-1}$ can evolve into J1627 in a timescale less than $10^{5}~\rm yr$. The models with
magnetar + fallback disk can account for the spin period, period derivative, and X-ray luminosity of J1627.

Because of a small upper limit of the period derivative, the origin of another radio transient J1839 remains mysterious. Our simulations show that the NS + fallback disk model can
account for the observed period and period derivative of J1839, while the required timescale is $\sim10^{7}~\rm yr$, which is much longer than the possible active timescale ($\sim10^{5}~\rm yr$) of a fallback disk. Therefore, we propose that J1839 may be in the inactive stage of a fallback disk similar to \cite{genc22}. Taking an active timescale of $10^{5}$ yr, the propeller torque of the fallback disk exerted on an NS can spin it down to a period of 1200 s in the quasi-spin-equilibrium stage. After the fallback disk becomes inactive, the NS can evolve to the current period and period derivative of J1839 in the second ejector phase. Those NSs with a magnetic field of $B=(2-6)\times10^{14}~\rm G$ and a fallback disk with an initial mass-accretion rate of
$\dot{M}_{0}\sim10^{24}-10^{26}~\rm g\,s^{-1}$ are the possible progenitors of J1839, which can evolve into J1839 in a timescale less than its characteristic age.

Our simulations also show that the evolutionary fates of NSs in the $\dot{P}-P$ diagram are very sensitive to the two input parameters including $B$ and $\dot{M}_{0}$. The NS with a weak magnetic field or a low mass-accretion rate naturally evolves toward a normal pulsar. The strong magnetic field NS + fallback disk model can interpret two ultra-long period radio transients and those long-period pulsars. It is worth emphasizing that ultra-long period radio transient J1627 with a high $\dot{P}$ ought to be in the quasi-spin equilibrium stage, while the transient J1839 and two long-period pulsars evolve to the second ejector phase at present.
Certainly, there are many uncertainties in the nature of J1839. A relatively accurate $\dot{P}$ may provide a reliable constraint whether or not it is a radio magnetar. More multiwave observations can also help us to unveil the mysterious nature of J1839.

\acknowledgments {We are extremely grateful to the anonymous referee for
helpful comments that improved this manuscript. We thank H. Tong, and K. Qin for
helpful discussions. This work was partly supported by the National Natural Science Foundation of China
(under grant Nos. 12273014 and 12203051), and the Natural Science Foundation
(under grant No. ZR2021MA013) of Shandong Province.}

\end{document}